\documentclass[nohyper,12pt,letterpaper]{JHEP3}

\usepackage{epsfig}

\def \eqn#1#2{\begin{equation}#2\label{#1}\end{equation}}

\title{Heretics of the False Vacuum:
Gravitational Effects On and Of Vacuum Decay. 2.}

\author{T.\,Banks\\

  SCIPP, University of California, Santa Cruz, CA 95064\\
{\it and}\\
NHETC, Rutgers University, Piscataway, NJ 08854\\

  E-mail: \email{banks@scipp.ucsc.edu}}

\abstract{This paper reexamines the question of vacuum decay in
theories of quantum gravity.  In particular it suggests that decay
into stable flat or AdS vacua, never occurs. Instead, vacuum decay
occurs, if at all, into a cosmological spacetime.  If the latter
has negative cosmological constant, it generically undergoes a Big
Crunch, which suggests that the whole picture is inconsistent. The
question of decay of de Sitter space must be very carefully
defined.  }


\preprint{\hepth{0211}\\RUNHETC-2002-44\\SCIPP-02/28}

\begin{document}




\section{\bf Introduction}

{\it The possibility that we are living in a false vacuum has never
been a cheering one to contemplate.  Vacuum decay is the ultimate
ecological catastrophe; in a new vacuum there are new constants of
nature; after vacuum decay, not only is life as we know it impossible,
so is chemistry as we know it.  However, one could always draw stoic
comfort from the possibility that perhaps in the course of time the
new vacuum would sustain, if not life as we know it, at least some
structures capable of knowing joy.  This possibility has now been
eliminated.  {\rm S. Coleman, F. De Luccia,1980}
}
\vskip.2in

In a series of beautiful papers written in the 1970s and 80s,
Coleman, with Callan and DeLuccia \cite{cdl} applied the instanton
methods invented for dealing with semiclassical decay in
statistical mechanics\cite{langer} and multivariable quantum
mechanics\cite{bbwlipsaclay} to the decay of a metastable vacuum
state in quantum field theory.  The present work, which takes part
of its name from the last paper in this series, is meant as a
critical examination of what these results might mean in a full
fledged theory of quantum gravity.

It is important to recognize the historical context in which the
Coleman DeLuccia paper was written.  Quantum and statistical
mechanics, including quantum field theory, primarily deal with
isolated subsystems of the universe. The Poincare invariant vacuum
state of quantum field theory, is a fiction, representing our
conviction that spacetime is locally approximately flat and that
the phenomena we are investigating are not terribly affected by
the global structure of the universe.  Coleman and DeLuccia
realized that this could not be true of a truly cosmological
application of the idea of false vacuum decay\footnote{I use the
word cosmological to mean a discussion of the state of the entire
universe.  In much of the literature on inflationary cosmology,
the part of the universe that we see is considered to be only a
small subsystem of a much larger metauniverse.   I will be
cautious, and express reservations about whether or not the
current discussion applies to the use of vacuum decay calculations
in an inflationary context. The question cannot be answered
without understanding the global structure of inflationary
cosmology.  It is inextricably intertwined with the discussion of
whether there had to be a pre-inflationary state, or some kind of
self-reproducing structure.}.   I will attempt to show that their
results, and extensions of them constructed here, suggest that
vacuum decay does not occur in many situations in quantum gravity.
In many situations in which it may occur, I will argue that it
cannot be thought of as the decay of one maximally symmetric
spacetime into another.

Much of what I have to say is simply reemphasis and
reinterpretation of the results of Coleman and DeLuccia.  Those
authors emphasized the use of the thin wall approximation to
vacuum decay, because it gave very explicit and calculable
expressions for decay amplitudes.  This approximation is however
misleading in one important respect.  It neglects the variation in
the scalar field (or fields) in most of spacetime and replaces it
by a discontinuous jump between constant values.  This is
sometimes a good numerical approximation for the decay amplitude,
but it gives the mistaken impression that the spacetime, both in
and outside of the bubble, is well approximated by a maximally
symmetric spacetime.  Coleman and DeLucia did not say this, and in
fact, emphasize that in the case of decay into a spacetime with
negative cosmological constant one rarely gets AdS space, but
rather a Big Crunch spacetime.  The actual result of this
calculation has been ignored in most later discussions of the
vacuum decay phenomenon\footnote{This is despite the fact that it
was the occasion for one of the most amusing pieces of rhetoric in
the Coleman canon.}.

The first conceptual point that I want to make about vacuum decay
is an elementary one.  The idea of decay of a metastable state,
presupposes the existence of a stable one, of which the metastable
state is an excitation.  In most (formal and informal) discussions
of vacuum decay one supposes this to be the maximally symmetric
spacetime that is found inside the bubble in the thin wall
approximation.  In the next section I will argue that the actual
results of Coleman and DeLuccia, as well as elementary
considerations about the formation of the metastable state
starting from a maximally symmetric stable vacuum, give the lie to
this supposition.  Vacuuum decay occurs, if at all, into open
F(riedmann)-R(obertson)-W(alker) cosmologies, whose global
geometry is very different from that of a maximally symmetric
spacetime.  The Big Crunch singularity of the CDL solution is, I
believe, an argument that decays into spaces of negative
cosmological constant do not occur in consistent theories of
quantum gravity.

Section 3 is devoted to the decay of dS spacetimes.  If dS
spacetime can really be thought of as arising from a quantum
theory with a fixed finite number of states\cite{tbfollywilly}
then the idea that it could decay into any sort of open universe
violates the rules of quantum mechanics.
What is really at issue is the question
of whether some open universe model (with an infinite number of
states) could contain a metastable subset of states which has
observables similar to those in a dS universe. Similarly, we could
ask whether the quantum theory of a dS space with one value of
the cosmological constant, might contain a factor space of states
which behaved for some time like a dS space of larger cosmological
constant.  There are thorny
conceptual problems to be solved here, as well as technical
problems with the definition of the instanton solutions, which
might represent the decay.  I will address these questions, and
the reader must judge whether I have resolved them.  The picture
I have been led to suggests that dS to dS transitions may occur
in a thermal manner.  The interpretation of the relevant instanton
is highly observer dependent, and various measurement theory issues
must be dealt with. Decays of dS space into an open, matter dominated
FRW universe with vanishing cosmological constant may also occur, if
one can establish a consistent theory of gravity in the FRW background.
This would require an understanding of the Big Bang singularity.
For some values of parameters in the Lagrangian, even transitions
to negative cosmological constant Big Crunch universes might make
sense.  For these parameter values, entropy bounds do not make the
assumed transition paradoxical.  I should emphasize that none of
these considerations prove that dS decay really occurs.  They simply
show that there is nothing in the semiclassical arguments
nor in general principles like holography and unitarity that
would prevent them from occurring.  To really demonstrate the
existence of these transitions, one must construct the quantum
theory of the putative stable ground state to which the dS space
decays.

In section 4 I will argue that the results of this paper apply
also to the membrane nucleation process discussed by Brown and
Teitelboim\cite{bt}, which has been used as a mechanism for
relaxing the cosmological constant\cite{relax}.

In the Conclusions I will discuss what these results mean for
various concepts in string theory, in particular the notion of an
effective potential.  I wish however to pause here to make clear
the spirit in which this paper was written.  Despite years of
work, we still do not have a clear and complete nonperturbative
definition of a quantum theory of gravity.  Matrix
Theory\cite{bfss} and AdS/CFT\cite{maldagkpw} give us definitions
in certain restricted circumstances.  The former, defined in light
cone gauge, is not suitable for the discussion of vacuum decay.
Its very formulation presumes the existence of null Killing
vectors that do not exist in the spacetimes into which flat space
is hypothesized to decay. While this by itself may be suggestive
, it is hard to argue that the
problem is with vacuum decay rather than with the light cone
formalism.

I will use some AdS/CFT wisdom to throw light on the problem of
vacuum decay, but for the most part my arguments will be, in the
spirit of CDL, reliant on properties of classical G(eneral)
R(elativity).  Fischler and I have argued elsewhere\cite{tbf} that
this may be one of our most reliable guides to the structure of
the quantum theory.

Finally, I would like to note that, as is the case for most {\it
Hollywood} sequels, the director of the original has long since
severed all connection with the project\footnote{He won't even
collect residuals.} .  As a consequence, although the present
paper may contain some grains of truth that were missed by CDL, it
is sure to be less entertaining.  The present author can only
apologize to his readers for this, and urge them to go back and
really read the work of the Old Master.

\section{\bf The real Coleman and De Luccia}

I will begin by summarizing the work of Coleman and collaborators.
However, it is convenient to first place it in a modern context.
In the 70s and early 80s, Higgs fields were the paradigm for the
scalars which occur in vacuum decay, and one was careful to try to
work in situations where gravitational corrections to local
physics were irrelevant.  Thus, it is assumed, and supposedly self
consistently verified, that gravitational corrections to instanton
amplitudes are small, and gravitational effects only become
important for the large classical bubbles which result from the
decay.

In the modern context, the most likely candidate scalars are {\it
moduli fields}.  Thus, we imagine a compactification of M-theory to $d\geq 4$
noncompact\footnote{For the moment, we will include dS space under
this rubric, though its spatial sections and Euclidean
continuation are compact.  We will soon see that this is very
important.} dimensions, and study the effective field equations of
approximate moduli\cite{tbcosmo}. Since it makes no difference to
most of our considerations, we will take $d=4$ for notational
simplicity. This requires us to be in a region of moduli space
where \lq\lq SUSY is broken by a small amount \rq\rq .  It is not
at all clear to me in what situations it makes sense to use the
phrase in quotes.  To fix ideas, imagine a scenario in which some
combination of asymmetric orbifolding, flux stabilization and/or
timelike linear dilatons in supercritical string theory have
convinced us that it makes sense to write down an effective
potential for ``approximate moduli'', with metastable minima.
Generically, we will get a
Lagrangian of the form:

\eqn{lag}{- M_P^2 \sqrt{-g} R - {1\over 2} G_{ij} (\phi / M_P )
\nabla \phi^i \nabla\phi^j - M^4 {\cal V}(\phi / M_P )}

Here $M_P$ is the four dimensional Planck mass, $M \ll M_P$, and
is supposed to be so because of the large fluxes, large number of
extra dimensions {\it etc.}.   In principle, there could also be
fields ({\it e.g.} boundary moduli in Horava-Witten
compactifications) whose natural scale of variation was $M$ rather
than $M_P$.  This would introduce small dimensionless parameters
into ${\cal V}$, as would inclusion of standard model Higgs fields (which
might be such boundary moduli) and
the like.

It is easy to see that a Weyl transformation, combined with
passage to the dimensionless variables $\phi^i / M_P$ removes all
trace of $M$ and $M_P$ from the equations of motion, and gives an
action proportional to $({M_P \over M})^4 $.  The dimensionful
scale of variation of all fields is $M_P / M^2 \gg 1 / M \gg 1/M_P
$. Thus, the hypothesis that $M$ is small in Planck units,
justifies the semiclassical approximation, as well as the neglect
of higher derivatives in the effective action.  There is, in
general, no excuse for further approximation, though the small
dimensionless parameters referred to above might provide one.

The solutions that we consider will all be four dimensional
spacetimes with a three dimensional maximally symmetric subspace.
Thus the metric will have the form \eqn{met}{ds^2 = \pm dz^2 +
\rho (z) ^2 d\Omega^2 ,} where the second term is the metric of a
manifold of a Lorentzian or Euclidean manifold with
$SO(4)$ or $SO(3,1)$ isometry group.  When
$z$ is timelike, we will call it $t$.  The scalar fields are
functions of $z$ only.   The solution thus defines a curve in the
space of fields, $\phi^i$ .  For any such curve, define the path
length $P$, by $P = \int dz \sqrt{G_{ij} d\phi^i /dz d\phi^j
/dz}$.  Then the scalar kinetic term in the Lagrangian will be
proportional to $(dP /dz)^2 $.   The potential term, followed
along the curve, will be some function $V(P)$ (a different function
for each curve in field space).  The variational
problem can thus be split in two: find stationary points of the
action for curves in field space with the constraint of fixed path
length, and then solve the time dependent problem of how the path
length varies with $z$ .   The latter is, up to notational
changes, the one field problem studied by Coleman and De Luccia. The
former is independent of gravity and depends only on the geometry
of field space and the shape of the potential on it.   The price
that we pay for this simplification is that the typical path
followed between two specified points in field space might be very
long, and the $V(P)$ might be a very complicated function with
many minima and maxima.  For example, the typical metrics found on
moduli space in the semiclassical approximation have chaotic
geodesics.  We must also worry about the fact that the
transformation to path length becomes singular when the field
space velocity vector, $d\phi^i /dz$ vanishes.  For the standard
CDL vacuum decay solutions this only happens at the end of the
trajectory, and causes no problem.  We will see that in the
analysis of dS decay, solutions with vanishing velocity are more
common. For the most part, we will ignore this complication, and
draw pictures of the simple double minimum potential used by CDL,
but we will comment on the more complicated generic situation at
various points.

Now let us recall the strategy of CDL.  We will begin by making
the assumption that the unstable vacuum has a negative
cosmological constant, and then examine the zero cosmological
constant limit.  The more subtle case of a positive cosmological
constant will be reserved for section 3.   The instanton is a
noncompact Euclidean geometry with metric \eqn{negmet}{ds^2 = dz^2
+ \rho (z)^2 d\Omega_3^2} which is a three sphere bundle over a
noncompact interval , $z \in [0,\infty ]$.  If we define, $U = -
V$, the Euclidean equations are

\eqn{euceqa}{(\rho^{\prime})^2 = 1
+ {\rho^2 \over 3} E}

\eqn{euceqb}{E = {{P^{\prime}}^2 \over 2} + U}

\eqn{euceqc}{{P^{\prime}}^{\prime} + 3({\rho^{\prime} \over \rho})P^{\prime}
+ U_P = 0.}

These are the Newtonian equations of motion for a particle moving
in the potential $U$ with a dynamically determined friction (or
anti-friction if $\rho^{\prime}$ is negative)\footnote{They are
also the equations for a scalar field in a negatively curved FRW
universe with potential $U$.  I am eschewing this interpretation
to avoid confusion with the real FRW cosmologies, which arise
after analytic continuation of the instanton.}.  The boundary
conditions are that as $z \rightarrow \infty$ the geometry
approaches that of Euclidean AdS and $P$ approaches $P_f$,
the false minimum of the potential
${V}$.   In fact, we should also require that near $z =
\infty$, $P$ is a normalizable solution of the linearized AdS wave
equation, since the instanton must represent an allowed
fluctuation in AdS quantum gravity.    The other boundary
condition is that $\rho$ must vanish at the point we have
conventionally chosen to call zero (the translation symmetry of
the equations is a residual gauge symmetry, and we fix it by this
choice).  The equations are consistent and nonsingular only if
$P^{\prime}$ also vanishes at this point.

 Geometrically, this is the requirement that the
Euclidean manifold be smooth and have no other boundary than the
one at infinity.  From the point of view of instanton dynamics,
the point $\rho =0$ will be the tip of the light cone to which the
bubble of true vacuum asymptotes (using the language of the thin
wall approximation).   We obtain a Lorentzian manifold by
analytically continuing the $3$ sphere to a $2+1$ dimensional dS
space, obtaining an asymptotically (Lorentzian) AdS manifold,
written as a dS fibration.  The dS fibration becomes singular when
$\rho$ vanishes, but the manifold is smooth.  We go to new
coordinates by doing a double analytic continuation in which the
dS space is continued to a Euclidean manifold with constant
negative curvature and $\rho (z) \rightarrow i a(i t)$.  $t$ is
then timelike and can be interpreted as the cosmic time of a
negatively curved expanding\footnote{Since $\rho$ starts from
zero, it must increase.} FRW universe, with equations of motion:

 \eqn{loreqa}{({\dot{\rho}})^2 = 1
+ {\rho^2 \over 3} E}

\eqn{loreqb}{E = {{\dot{P}}^2 \over 2} + V}

\eqn{loreqc}{\dot{\dot{P}} + 3{\dot{\rho} \over \rho}\dot{P} + V_P = 0,}
where dots denote $t$ derivatives.  In order to keep the
notation simple, we have used the same letter to denote the
Lorentzian and Euclidean energies.  They are not the same
quantity, but we hope this will not cause confusion.
 This spacetime seems to have a
big bang singularity at $t = 0$, but since $\dot{P} = 0$, this is
no more dangerous than the apparent singularity in the FRW
coordinates of AdS space.   The space is not however AdS. We will
see in a minute that $P(0)$ cannot be sitting exactly at the true
vacuum.  It is determined instead by the conditions on the
instanton at infinity.  This accords with our usual ideas of
tunneling: we tunnel not precisely to the true classical ground
state, but to some point on the potential in its basin of
attraction.

As the universe inside the bubble expands, the energy $E$
decreases and becomes negative.  Thus $\dot{\rho} / \rho$
decreases and eventually goes to zero.  If $\dot{P}$ were exactly
equal to zero at this point, the universe would stop expanding and
$P$ would oscillate for ever around its minimum.  This bizarre but
possible state of affairs will not happen generically because the
boundary conditions are all fixed at $t = 0$.  Instead, the
universe begins to recontract and eventually returns to $\rho =
0$.  If $\dot{\phi}$ happened to be exactly zero at this point, we
would have indeed reached the true AdS vacuum.   But again, this
is a highly nongeneric situation which could only be achieved by
fine tuning the potential.   The actual solution has a Big Crunch
Singularity for almost every potential. Thus, the conclusion of CDL is that
if the true minimum has a negative cosmological constant (as it
must if the false vacuum is AdS or flat) then the universe tunnels
not to AdS space, but to disaster.  We will reserve our comments
about the meaning of this result until we have examined the
instanton solution itself.

 \FIGURE{\epsfig{file=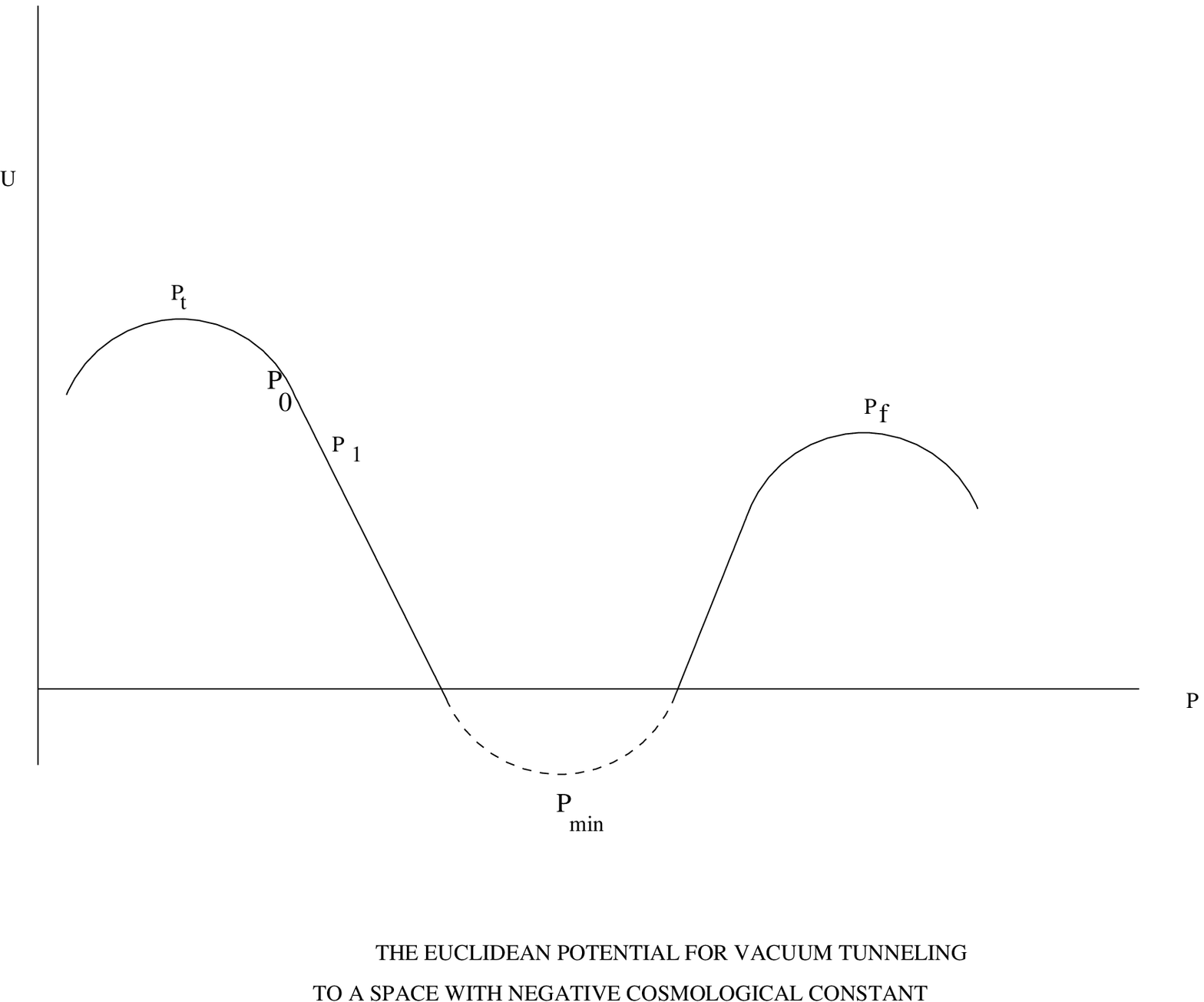}}

The potential $U$ is drawn in Fig. 1 . Note that we have drawn
the region near the minimum as a dashed curve, below zero.  If the false vacuum
energy were zero, this would be a necessity, but if it is negative
it could be that the minimum of $U$ is positive. The dashes are
supposed to indicate that this segment of the potential could be
above zero. Our particle starts with zero velocity, near the
higher maximum of $U$ at $z=0$.  It cannot start precisely at the
higher maximum, because the solution with those boundary
conditions remains at the higher maximum for all $z$ and does not
asymptote to the Euclidean AdS space at the false maximum.  Recall
that starting precisely at the true maximum was the only way to
save the system from the Big Crunch.

We now ask whether there are any solutions which have the right
asymptotics.  The parameter we have at our disposal to vary is the
initial position of the particle.  If there were no friction, we
would be guaranteed a solution.  Indeed, in that case energy is
conserved and the particle would reach zero velocity at a value
$P_{f}$ where the potential energy is the same as that at the
initial point. If $P_{f}$ is a stationary point of the
potential the transit time is infinite, as desired.

When we deal with a system with friction, such a conclusion is no
longer guaranteed.  We must surely start the particle out to the
left of the point $P_1$ whose energy equals $U(P_{f})$.  In
the case where gravity is neglected, Coleman argued that by
starting the system very close to the higher maximum one could
guarantee that the particle overshoots $P_{f}$ in finite
time.  Indeed, in that case the friction coefficient is just
$\propto 1/z$ and is always positive.  If we start with zero
velocity, close to the maximum, then we stay there until very
large $z$, friction is negligible, and when the particle finally
moves it overshoots.  When gravity is included, this strategy is
ineffective, and in fact counter productive.  By going to large
$z$, keeping $P$ near the maximum, we keep ${{{P^{\prime}}^2}\over 2} + U(P)$
approximately constant and the friction coefficient stays
approximately constant while $\rho$ gets large.  This is like the
slow roll regime of inflationary theories, where curvature is
inflated away.  But our goal is to get $P$ to its second maximum
as $\rho$ goes to infinity and we have not progressed towards that
goal.  We still have motion in a potential with a substantial
amount of friction and it is not clear if there will be enough
energy to get to the second maximum.

If $U \geq 0$ everywhere, depending on the details of the
potential, we may find that $P$ always oscillates down to the
minimum of $U$ as $\rho\rightarrow\infty$ .  This would define a
noncompact manifold, but would not be an instanton relevant to
false vacuum decay.  If $U$ is negative somewhere (as it must be
if the false vacuum is at zero energy) an even worse disaster may
occur. $\rho^{\prime}$ may go to zero at a finite value of $z$.
$\rho$ then turns around and begins to decrease, and we do not
even get a noncompact instanton of the correct topology.

We see that, depending on the details of the potential $U$, there
may or may not be an instanton that describes tunneling of a flat
or AdS vacuum into a spacetime of lower cosmological constant.
This is a generalization of a famous result of CDL.  In the thin
wall approximation, only one parameter characterizing the
potential appears in the final answer - the vacuum energy
difference. CDL show that when this parameter is too small,
tunneling does not occur.  Beyond the range of validity of the
thin wall approximation, the answer to this question depends on
the detailed structure of the potential.

We can also conclude from this analysis that in string
compactifications with many moduli the chance that there is an
instanton is likely to be smaller.  We showed that the effect of
the higher dimensional moduli space and the chaotic behavior of
its geodesics would be to make the distance in $P$ between the two
maxima of $U$ large, and to introduce subsidiary maxima and minima
into the effective potential.  It seems clear that this will
reduce the chance of the particle getting to the false vacuum
maximum of $U(P)$.

Given the CDL conclusion about the nature of the spacetime that
the instanton decays to, this is a welcome turn of events.  The
Big Crunch spacetime does not appear to be
a stable ground state of anything.   One might imagine a
nongeometrical description of the quantum physics of this system,
takes over at the Big Crunch.  However, according to the FSB
entropy bound\cite{fsb}, no observer in the spacetime can ever
discover more than a finite (and fairly modest) number of quantum
states.  On the other hand, experiments done in the false vacuum,
on times short compared to its decay time, can establish the
existence of a much larger number of states.   Thus, there seems
to be a paradox in imagining one could decay to the other.

The nonexistence of the instanton for large classes of models
provides us with a way of escaping this paradox.  We are led to
conjecture that in sensible theories of gravity in AdS or
Minkowski vacua, there are no instanton solutions.  This is of
course well known to be the case for supersymmetric vacuum states
\cite{weinbergcvetic}.  In the thin wall approximation, the CDL
bound on the vacuum energy difference necessary for vacuum decay,
coincides precisely with the BPS bound on domain wall tensions.
For two Susic vacua there is a static domain wall rather than
a vacuum decay bubble.  I suggest that no sensible theory of
quantum gravity will contain a CDL vacuum bubble describing the
decay of an asymptotically flat or AdS spacetime.

My conclusion from this analysis is that the semiclassical approximation
gives us no evidence that decays of asymptotically flat or AdS vacua
ever occur in sensible theories of gravity.  As far as I can see,
the only loophole in this argument is the possibility that supersymmetric
theories might provide us with fine tuned potentials that guaranteed that
the interior of the Lorentzian bubble was nonsingular.  Numerical analysis
of examples seems to show that this is not the case.

The AdS/CFT correspondence provides strong evidence that this interpretation
of the semiclassical results is correct.   According to the duality, the
cosmological constant in Planck units
is identified with a discrete parameter characterizing
the CFT, for example a power of the rank of a gauge group.  Smaller
cosmological constants correspond to CFT's whose high energy density of
states grows more rapidly at infinity.  The limit of flat space has
a density of states which grows more rapidly than an exponential.
From this point of view, it is difficult to understand how a system could
tunnel into another with a smaller density of states.  In non-gravitational
contexts, tunneling is a low energy phenomenon.  For energies above the
barrier, the true and false vacua become identical.   AdS/CFT shows us
without any doubt, that this is not true in quantum gravity.

If the low energy effective field equations gave unambiguous evidence for
tunneling in a regime in which their validity was not in question, we would
be faced with a paradox.  Fortunately, they do not.   Depending on the
potential, we find either no tunneling, or tunneling to a state where effective
field theory breaks down, and we seem to be faced with an entropy paradox.
My conjecture would be that for any potential that we reliably compute from
a genuine theory of quantum gravity, we will simply find that there are
no instantons describing decays of flat or AdS vacua.

\section{\bf dS dK}

\subsection{The generic prescription}

We now come to the question of the decay of dS space, about which
there has been controversy both in the past\cite{hmlinde} and
more recently\cite{mss} .  Indeed, it was the latter papers that
stimulated my renewed interest in this issue.  A very clear discussion
of many of the issues can be found in a paper by Rubakov\cite{rub}.
However, this author works in planar coordinates, which are ill suited
for noticing many of the points that I will discuss below.

Before beginning I want to make a couple of points clear.  First
of all, if we admit that dS space is described by quantum
mechanics with a fixed finite number of states, then it is obvious
that it can decay neither into AdS space, flat space, a
nonaccelerating FRW universe, nor a dS space with a smaller value
of the cosmological constant.  These systems all have larger
Hilbert spaces than the original dS space.  Whatever else quantum
gravity does, if it obeys the most general principles of quantum
mechanics, it cannot describe change of dimension of the Hilbert
space.  These remarks do not settle the issue, even if we believe
that ``real" dS space has a finite number of states.  Attempts to
find dS minima in controllable approximations to string theory
will almost inevitably find them as metastable minima of an
effective potential that goes to zero at infinity in moduli space.
It is perfectly conceivable that the infinite system defined by a
solution that stays near infinity in moduli space, could have a
finite dimensional metastable subsystem that behaves like dS space
for some period of time . One cannot use properties of finite
systems to criticize this proposal ({\it e.g.} Poincare
recurrences or a finite horizon) because the approximate dS space
has to be presumed to be in contact with the other degrees of
freedom of the system if one is not assuming from the start that
decay does not occur.   Thus, it makes sense to revisit the CDL
proposal for dS space in the context of hypothetical metastable
minima.  Note that contrary to the situation we encountered in
discussing the decay of flat or AdS spaces, there is no
paradoxical problem of trying to fit a large number of states into
a box that is too small for them.  In the decay of a noncompact
vacuum, the false vacuum has a spectrum of high energy black hole
excitations which is incompatible with any state of the system
with more negative cosmological constant.  In the decay of dS space,
the true vacuum will have more states than the false vacuum as long
as the true cosmological constant is nonnegative.

Another possible red herring in this discussion is the ``decay" of
dS space into a black hole\cite{gpbh}. Unlike the analogous decay
of hot flat space\cite{gpy} this is not really a decay.  Viewed
thru the lens of the static Hamiltonian, dS space is a thermal
system.  There is a finite probability for anything to be
nucleated, including black holes.  But unlike hot flat space,
there is a maximal size black hole in dS space, and (for large dS
radius in Planck units) it has much less entropy than the ``dS
vacuum". It decays back into the vacuum state.  Indeed, it is much
more probable to nucleate a black hole that is not at rest with
respect to the static observer, and falls back into the horizon
much more rapidly than it decays.  After this point it is
interpreted as one of the states of the "dS vacuum ensemble" by
the static observer.

The first really troubling argument about the decay of dS space
into flat space has to do with the inverse process.  As mentioned
in the introduction, decay into a stable state implies the
possibility of producing the metastable state as an excitation of
a stable one.  This possibility was studied long ago by Guth and
Farhi\cite{gf}.  They showed that any attempt to create a local
region of an asymptotically flat spacetime that was
in a metastable dS vacuum led instead to the creation of a
black hole.   Furthermore, a singularity separates the external
observer from the dS region, so even observers who jump into the
black hole cannot see it.  The mass of the black hole
produced is determined not just by the value of the potential at
the dS minimum, but also by the walls of the potential surrounding
it.   But as the dS minimum is made lower, we can also lower the
walls (and keep metastability), so in this limit, the black hole
mass is small.   There is then a contradiction between the number
of states of the system counted by the Bekenstein-Hawking entropy
of the black hole, and the number counted by the Gibbons-Hawking
entropy .

It is clear then, that one cannot create a metastable dS minimum
as an excitation of flat space.  Even if we wanted to resort to
black hole complementarity \cite{tHstu} to invoke a complementary
description of physics inside the black hole that might look like
dS space, the clash of the number of states imputed to the system
by the internal and external calculations, prevents us from doing
so. Complementary observers may use non-commuting Hamiltonians to
describe physics, but they agree on the Hilbert space dimension.

However, even this objection to metastable dS vacua may be
overcome.  Recall that CDL showed that the spacetime inside the
bubble is {\it not} a symmetric space, but rather an open FRW
cosmology.  In the case of negative cosmological constant this
leads to a disastrous Big Crunch, which we have chosen to regard
as a signal that the decay will not occur in a sensible theory of
quantum gravity\footnote{Actually, even this has to be rethought
in the dS case, as we will see below.}.  However, a dS space can
decay into a spacetime with non-negative cosmological constant,
and in this case the FRW cosmology has a perfectly smooth future.

So the real question we must ask is whether this FRW cosmology
defines a consistent set of asymptotic boundary conditions for
quantum gravity.   This is a thorny question, for any such
cosmology has a Big Bang at a finite cosmic time in the past. It
is conceivable that Big Bang cosmologies are acceptable.  In
\cite{bf}, Fischler and the author argued that the Big Bang
state should be thought of as one in which the universe was in a
tensor product state of some number of two state systems.  Each
two state system describes the physics in a single, smallest,
horizon volume.  There is nothing singular about the quantum
mechanics, though the geometrical description breaks down because
areas are of order Planck scale.

The question of whether a metastable dS state can be formed in
the throes of a hot Big Bang,
 was answered in the affirmative in Guth's original
inflationary models.  One may question the validity\footnote{I am
speaking here of mathematical validity.  These are certainly not
good phenomenological models of cosmology.} of those models and
the logic that led to them, but we do not really have a clear
answer on this point.  It is part of a general question of what
the initial conditions are for inflationary cosmology, and whether
the inflationary scenario is robust or fine tuned.   It will be
hard to answer these questions without a more complete quantum
theory of cosmology than we have at present.

 Thus, at the present time, there seems to
be no paradox in assuming the existence of a metastable dS state
that decays into an open FRW model with non-negative cosmological
constant\footnote{The reader may wonder why I harp on the
distinction between a late time open FRW model and asymptotically
flat space.  Although the local physics of these two spacetimes is
similar in the asymptotic future,
their spatial asymptotics are very different, and one has a Big Bang
singularity which the other does not have. I believe the
fundamental formulation of quantum theories of gravity depends
crucially on the asymptotic boundary conditions .}.

Such dS decay would seem to be particularly natural for the models
of \cite{mss}.  These models have a potential, which goes to zero
in the weak coupling region of moduli space.  There are
cosmological solutions, with negative spatial curvature in which
the moduli remain within the weak coupling regime.  It would seem
natural for the metastable dS minimum constructed by these authors
to decay into the state at infinity.
However, they claim that within the range of validity
of their approximations, no appropriate instanton exists.  The
current paper was begun in an attempt to further investigate this
claim.

 Thus, with
these preliminaries out of the way, we can turn to the technical
problem of constructing the CDL instanton beyond the thin wall
approximation. Here I find myself differing substantially from the
CDL analysis. CDL's calculational approach is based on the idea of
small deviations from flat space results, at least for the
computation of the probability for the bubble to materialize.
However, there is a qualitative difference between the Euclidean
AdS and flat spacetimes, on the one hand, and dS on the other.
Euclidean dS space is $S^4$, a compact manifold without boundary.
If we ask for a spacetime of the form \eqn{comp}{ds^2 = dz^2 +
\rho (z) d\Omega^2,} with the same topology, then it must have two
values of $z$ where $\rho$ and $P^{\prime}$ vanish.  Set $z=0$ at
the vanishing point closest to the true vacuum maximum of $U$, and
at a point $P_0$ to the right of it in Fig. 2 . As before, we
cannot take $P_0$ to be the maximum itself, because the solution
would not move from there.

 \FIGURE{\epsfig{file=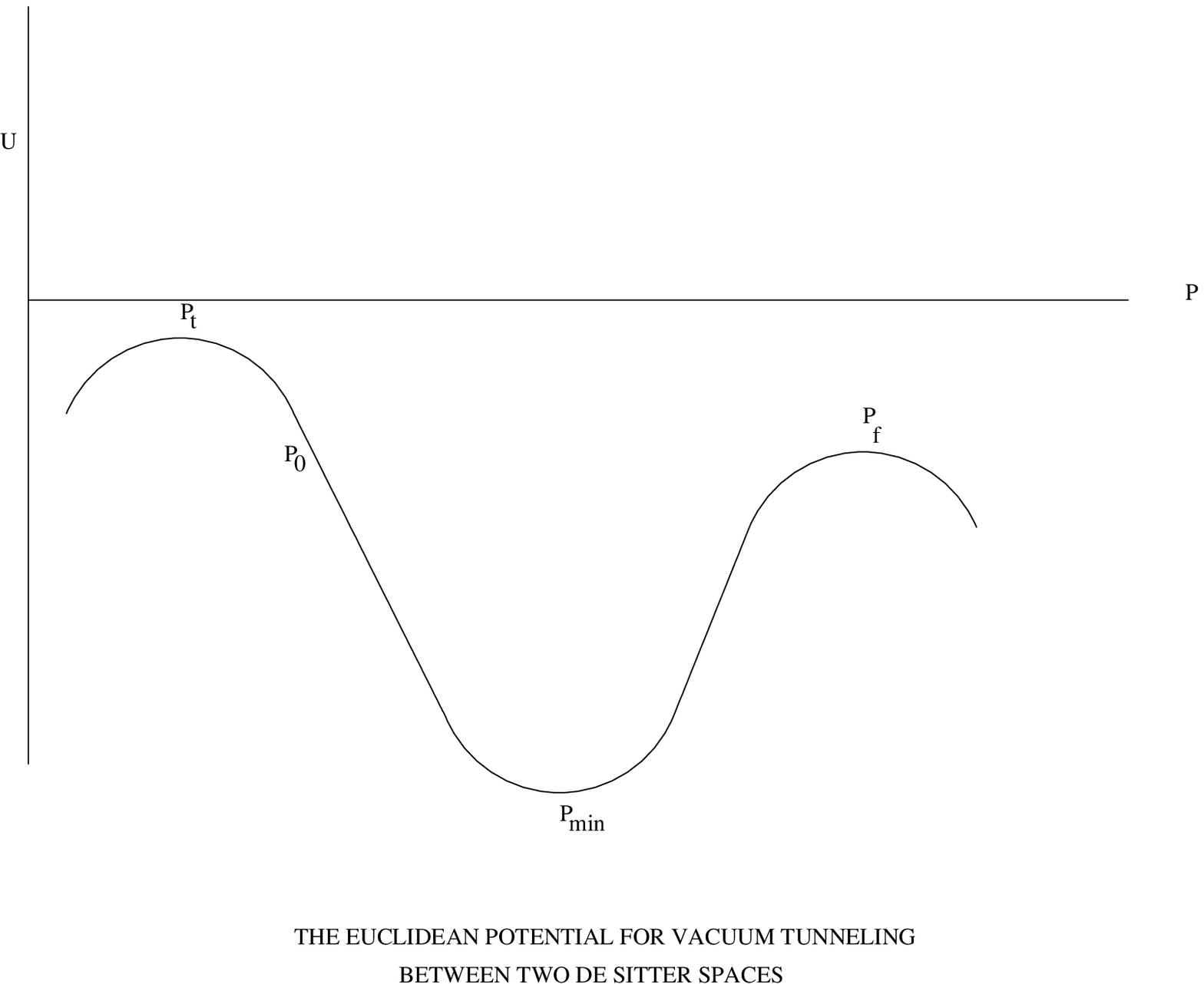}}

Let us consider the case where the true vacuum has non-negative
cosmological constant.  Then the potential $U$ is everywhere
negative.   In the Newtonian model system, the particle starts
with zero velocity at $P_0$ and falls to the right, slowed by
friction.  $\rho$ is increasing and the energy is decreasing. If
we choose $P_0$ to the right of $P_1$, the point where the energy
is equal to that at $P_f$, then the particle gets stuck in the
well and its energy goes negative.  Inevitably, $\rho^{\prime}$
goes to zero and changes sign.  $\rho$ begins to decrease, anti-friction
sets in and the energy increases.

It is clear that $\rho$ will get to zero at a finite value of $z$.
The question is, can we tune $P_0$ so that $P^{\prime}$ also
vanishes at that point?  The answer appears to be yes.  There is a
solution of the equations with constant $P=P_m$,  the minimum of
$U$.  If $P$ sits at this point, there is an oscillatory solution
for $\rho$, with a period $(- U_{min})^{- 1/2}$.   If we set
$P_0$ just to the left of this point $P$ will also oscillate, for
a while with a frequency given by the curvature of $U$ at its
minimum.  By tuning the initial value of $P_0$ we can make a zero
of $P^{\prime} $ coincide with a zero of $\rho$.  So we have found
a compact Euclidean instanton for this system, but it is not clear
what its relation is to vacuum tunneling.  The Euclidean solution
never visits either the true or the false vacuum.  Furthermore, in
the thin wall approximation $P$ never oscillates, so our
oscillating solution looks nothing like the thin wall solution.

What does it describe?  If we follow the CDL prescription, we must
analytically continue some coordinate of the sphere, to Lorentzian
signature in the region between the two zeros of $\rho$.  This
gives a Lorentzian four manifold foliated by $dS_3$ cross sections.
The scalar fields vary in a complicated manner in the spatial $z$
coordinate.   It is this peculiar manifold that we should think of
as the state to which (either the true or false) vacuum tunnels.
There are then {\it two} zeroes of $\rho$
across which we must analytically continue.  This means that this
instanton always corresponds to the nucleation of {\it two}
bubbles, which then grow at a rate approaching the speed of light.
If one of the zeroes of $\rho$ occurs at a value of $P$ that is in
the basin of attraction (for the true potential $V \equiv - U$) of
the false vacuum, then the Lorentzian continuation of the
instanton in this bubble, will relax back to the point $P_0$.

The geometry is initially spatially
curved, but as the field settles in to $P_0$ it begins to inflate.
Locally, the geometry then approaches dS space in planar coordinates, and
has a finite area cosmological horizon.  Thus, asymptotically,
every local observer finds herself in empty dS space.  {\it Unlike the
case of asymptotically infinite spaces, the existence of cosmological
horizons implies that we cannot describe observations that would
distinguish between this spacetime and dS space.}  Thus, I will
interpret the asymptotic behavior as relaxation to the dS vacuum state.

Inside the true vacuum bubble, things are as they were for vacuum
decay in the CDL analysis.  One asymptotes to the "true vacuum" if
it is dS space, to a negatively curved FRW if $\Lambda_{true} = 0$
and does not asymptote at all (there is a Big Crunch) if the true
cosmological constant is negative.

In fact there will be many
solutions of this sort\footnote{This is probably the place to
remind the reader that our parametrization of the multifield
problem in terms of path length, will be singular for the
oscillating solutions we are describing here.  The reader may if
he wishes, regard $P$ as a single scalar field from this point on,
in which case the discussion needs no modification.  As far as I
can see, the multifield case does not introduce any new features
except a much larger family of solutions.}.  Consider the
variation of the solution as we vary the value of $P_0$ between
the true and false maxima.  There might be some solutions starting
near the true maximum, which overshoot the false maximum.  These
will not lead to compact manifolds (unless there are other minima
of $U$).  Since the derivative of $P$ does not vanish when $\rho$ goes
to zero for the second time, such solutions are singular.
However, there is certainly a point $P_s$ somewhere to
the left of $P_1$, for which the solution does not overshoot. Such
a solution will remain trapped in the basin of attraction of $P_m$
until $\rho^{\prime} = 0$ and $\rho$ begins to decrease.  Then it
may oscillate a few more times and try to escape over one or the
other of the maxima.  By tuning $P_0$ we can make the last
oscillation of $P$, a point where $P^{\prime} =0$, coincide with
the point where $\rho$ vanishes. There will be a discrete set of
such solutions. Each will correspond to a two bubble space time,
after analytic continuation. The interiors of the two bubbles can
be relaxing to either the same vacuum (true or false) or different
ones, depending on the choice of $P_0$.  There is in fact an
accumulation point at $P_m$ of initial conditions which give a
compact geometry.  Thus there is a discrete infinite set of smooth
compact instanton solutions.  The situation is somewhat
reminiscent of the discrete infinite set of Einstein metrics on
higher dimensional spheres\cite{Bohm?}\footnote{I would like to
thank G. Horowitz for bringing this work to my attention.}. The
solution with $P_0 = P_m$ is the Hawking-Moss
instanton\cite{hawk}.  However, the Lorentzian continuation of
this solution does not look like either one or two vacuum bubbles.
It is simply a dS space with cosmological constant equal to the
value of $V$ at its maximum and $P$ sitting at the maximum.  It
clearly has a classical instability.   Nonetheless, the action of
the infinite set of two bubble instantons approaches that of the
Hawking-Moss instanton as the initial point approaches $P_m$.

The Euclidean action of a general solution of the instanton
equations is given by
\eqn{eucact}{S_E =- 4\pi^2 \int dz [\rho^3 U + 3\rho].}
It takes on the largest possible negative value for solutions which
stay for a long time in regions where the absolute value
of $U$ is small\footnote{Tunneling probabilities are given by
exponentials of differences between the action of two solutions.
and are always less than one.}.
In particular the solutions which accumulate near the Hawking-Moss
instanton give subleading contributions to the tunneling probability.
There is no reliable way to take them into account.   It is intuitively
plausible, but I have not been able to prove in general, that the
dominant contribution always comes from the instanton with the smallest
number of oscillations.  Like the flat space bounce, this solution makes
one traverse from the vicinity of the true vacuum to that of the false
one.  The Lorentzian history of the decay consists of the
nucleation of two bubbles in a compact ellipsoidal three space, with
spatial metric
\eqn{spatmet}{ds_3^2 = dz^2 + \rho^2 (z) d\Omega_2^2 ,}
where $d\Omega_2^2$ is the metric of a two sphere.  The poles of the ellipsoid
are the nucleation points of the bubbles.   The bubble walls accelerate
to the speed of light, but the space between them expands more rapidly.  The
two bubbles remain forever outside each other's horizon.   The interior of
each bubble has spatial sections which are homogeneous spaces of constant
negative curvature.  If both the true and false vacuum energy are positive,
then a local observer inside each bubble will rapidly find himself in
an environment resembling empty dS space with the appropriate value of
the cosmological constant.  If the true vacuum energy is zero, the observer
in one of the bubbles will
experience (asymptotically) a negatively curved, matter dominated
open FRW universe.  If the true vacuum energy is negative, this observer
will instead experience a Big Crunch.     Note that we cannot make as strong
a statement about the implausibility of Big Crunch solutions in the dS
context, if we believe that there is any sense in which the metastable
dS space has a finite number of states.  For some potentials, the number
of states accessible to observers in the Big Crunch universe may be
larger than those available to the false vacuum dS observer.

An important technical point, which I have not been able to resolve
is a calculation of the number of negative eigenvalues for this instanton
solution.  Coleman's general argument\cite{sidneg} for a single negative
mode for instantons in quantum mechanics and field theory conspicuously omits
discussion of the de Sitter case.  The issue has to do with gauge invariance.
The action

\eqn{act}{S = \int dz [\rho^3 ((P^{\prime})^2 /2 - U) - 3( \rho (\rho^{\prime})^2 - \rho)]}
for the equations of motion we have been studying, is a gauge fixed version
of the reparametrization invariant action
\eqn{tract}{S = \int dz e(z) [\rho^3 ({(P^{\prime})^2 \over 2 e^2} - U) - 3( \rho {(\rho^{\prime})^2 \over e^2}  - \rho)]}
Indeed, the Friedmann equation is properly viewed as the equation obtained
by varying the einbein $e(z)$.   One can gauge fix to constant $e$, and then
incorporate the constant value of $e$ in the parameter length, $L$, of the $z$
interval.  This is why all values of $L$ are allowed.   The question now is
how to properly count the fluctuation modes, including $L$.  For example,
for fixed $L$, the fluctuation corresponding to an infinitesimal $z$
translation of the instanton, which is formally a zero mode, does not
satisfy the boundary conditions.  So the usual argument for the existence
of {\it at least} one negative mode, does not hold.  A related problem
is that the full system of equations is not a Sturm-Liouville system.
One must do a complete gauge fixing and eliminate all but physical modes
(which is awkward ) to get a proper answer to these questions.

It is interesting to compare these tunneling arguments for dS space
with finite temperature tunneling in quantum field theory.  At finite
temperature, Euclidean time is also compact.  For low temperature there
are an infinite set of periodic instantons, which must all be taken into
account.  The instanton with $N$ oscillations, gives the term of order
$T^{-N}$ in the expansion of the exponential of the free energy.  Half
of these instantons have an even number of negative modes and half have
an odd number.  This corresponds to the expansion of $e^{i\Gamma}$ where
$\Gamma$ is the width of the metastable state.   For temperatures which are
not low, we do not have to worry about large inverse powers of $T$ and
only the lowest action instanton contributes to the calculation. Our
dS calculation for generic values of parameters in the potential is
similar to the finite temperature case at temperatures of order one.

One might similarly guess that half of our dS instantons have an even and
half an odd number of negative modes.   In particular, my guess is that the
lowest action instanton, which does not oscillate at all,  has precisely one.
I hope to report a proof of this conjecture, as well as the conjecture that
the lowest action solution does not oscillate, in a future publication.

Let us return to the interpretation of the instanton solution. The Lorentzian
continuation of the instanton is a spacetime with observer dependent
horizons.  To see this, introduce static coordinates in the dS fibers of
that portion of the manifold between the two zeros of $\rho (z)$.  The metric
is
\eqn{statmet}{ds^2 = dz^2 + \rho^2 (z) ((1- r^2) d\tau^2 +
{dr^2 \over {1 - r^2}} + r^2 d\Omega^2) ,}
where $-\infty \leq \tau \leq \infty$ and $d\Omega^2$ is the metric on
a unit $d-3$ sphere.   These are the natural coordinates for
a timelike observer following a geodesic
with fixed $z$ at the point where $\rho^{\prime}=0$,
and fixed position on the $d-2$ sphere.  She sees a static
ellipsoidal geometry.   The Penrose diagram of the Lorentzian bubble space-time
(Fig. 3) shows that a finite volume of the phase space of timelike
geodesics, correspond to observers for whom both bubble walls are outside
the cosmological horizon.
 \FIGURE{\epsfig{file=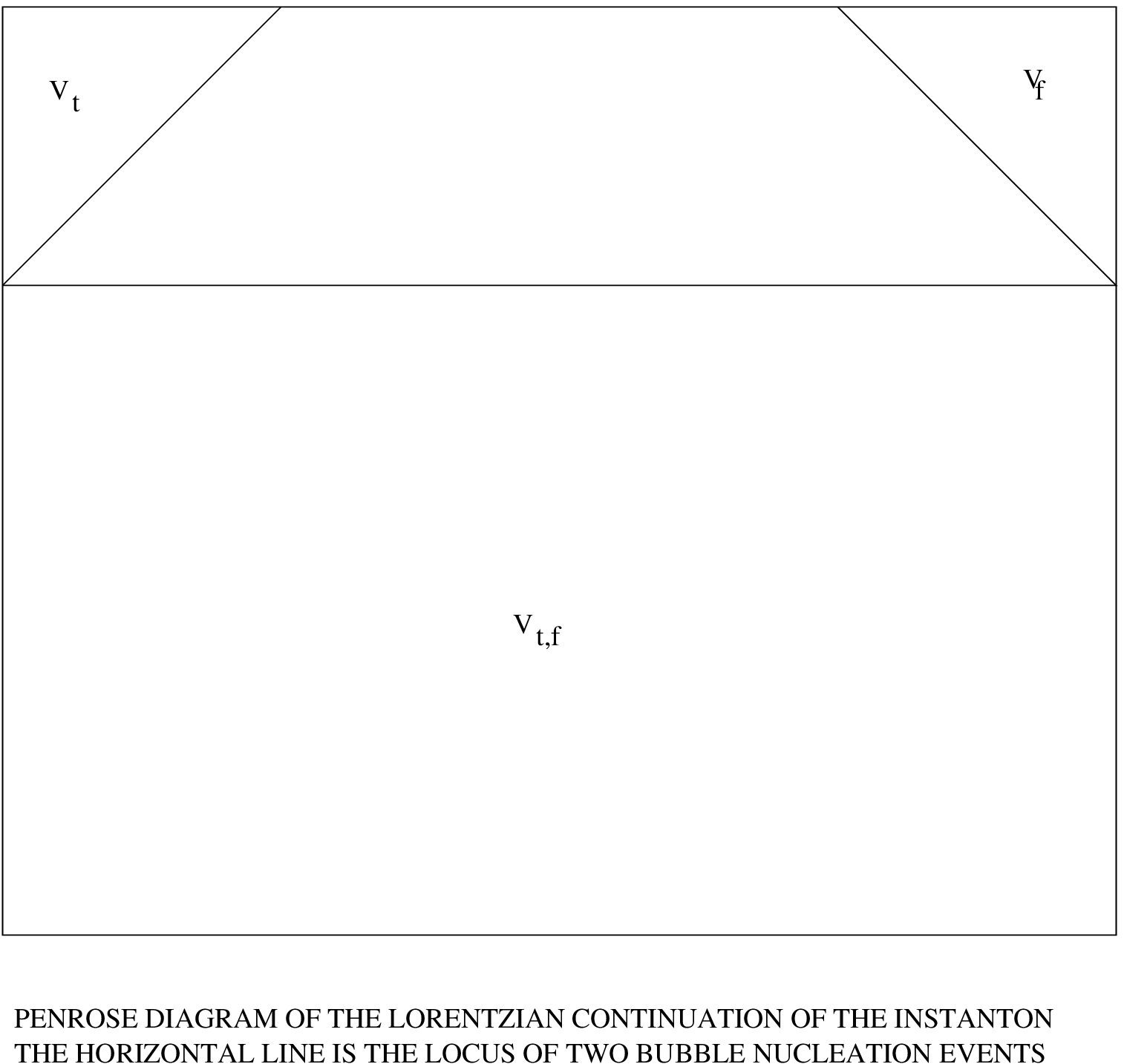}}

At the present time, quantum measurement theory depends (in my opinion)
on the description of the measuring apparatus by an approximate local field
theory.   The device should thus be thought of as a localized low energy
observer, following a timelike trajectory in spacetime.  Thus, many
devices will not register the existence of the vacuum bubbles.

The correct state of the field theory describing a device using the
static coordinate system, should be a
local excitation of the Euclidean vacuum, whose Green's functions are
defined by analytic continuation from the Euclidean section.  Thus, these
static observers will see a thermal state, with $z$ dependent temperature.
The temperature goes to infinity at the zeroes of $\rho$, as well as at
$r = 1$, for all $z$.  The temperature at the observer's position is finite,
and, as long
as there are no large or small dimensionless parameters in $U(P)$, it
is of order $M^2 / M_P$.  Again it is apparent that
these observers see no vacuum decay.
This observation is a restatement of the problem of the failure of
the true vacuum bubbles to percolate through the universe in old inflation
models\cite{guthweinb}.  The difference is that, in old inflation models
the false vacuum was a temporary feature of the effective potential
caused by finite temperature effects.  Here we are speaking of a global
dS vacuum.

There are of course local
observers in this spacetime who experience the interior
of the two expanding bubbles.  However, they are causally disconnected
from each other, as well as from the static observers of the previous
paragraph.  There are three kinds of freely falling
timelike observers in
this spacetime.  Many timelike geodesics miss both vacuum bubble
light cones.  Observers following these trajectories can be viewed as
moving observers in a static, thermally
excited spactime.   Those observers on the submanifold with
$\rho^{\prime} (z) =0$ are completely static.
Observers who pass through the
the vacuum bubble light cones, penetrate into the vacuum bubbles.
They experience a dS vacuum (with either true or false values of the
cosmological constant), which is excited by a tunneling event to a
non-vacuum value of the homogeneous mode of the scalar field.  The resulting
Bose condensate is then diluted by inflation and the local observer sees
the dS vacuum again.

The picture of vacuum tunneling event is thus much more complicated
in the dS case than in asymptotically infinite spaces.  The spacetime before
the tunneling event is the false dS vacuum.  For local observers, this means
a thermal state of quantum fields in the false dS background.  After the
tunneling event, observers break up into three classes.  Some indeed see
an abrupt transition to a regime that locally asymptotes to the true vacuum
(which in the case of vanishing cosmological constant is a negatively curved
FRW universe rather than Minkowski space).
Some see a static ellipsoidal spacetime with position dependent temperature.
We will refer to these observers with the adjective {\it inhomogeneous}.
Others find themselves in an excited state of the false vacuum, which
quickly reverts to the false vacuum, because of inflation.

Note also, that since the manifold is compact, there is an equally good
interpretation of this instanton as an event occurring in the true vacuum.
The difference is only in the manifold we choose to glue in to the past
of the Lorentzian continuation of the instanton.  Since we are interpreting
this solution as a tunneling event and are consequently willing to
tolerate discontinuities in the classical evolution, this choice is arbitrary.

These results have an obvious thermal interpretation\cite{rublinde}. All
local observers in spacetimes with horizons, experience a temperature.
In both the false and the true vacuum, there is some thermal plus quantum
probability to nucleate the state of the fields described by the initial
conditions of the Lorentzian continuation of the instanton.   Although
the instanton is the same for the jump from true to false as for false
to true vacuum, the probability is different.  The log of the probability
is given by $(S_{F,T} - S_I)$, for the false and true vacuum decays
respectively.  Here $S_I$ is the instanton action and $S_{F,T}$ the actions
of the true and false dS spaces.
I would like to  interpret this as a statement of detailed
balance.   That is, in a system with thermal excitation we can have transitions
between the two states described by the stable and metastable minima of the
potential.  The total probability of transition is thus determined both
by an intrinsic matrix element, and by the number of final states.
The ratio of the two transition probabilities is the ratio of the number of
initial and final states.

W. Fischler\footnote{Private Communication} brought up the question of
whether the entropy measured by the static inhomogeneous observer in the
aftermath of the instanton event is always less than that of the true
vacuum dS space.  This would be required if we were to interpret the instanton
in terms of the quantum mechanics of the true dS vacuum.  Then the
full instanton, as seen by any observer, could consistently be viewed as an
excitation of the true vacuum state.  The principle of Cosmological
Complementarity\cite{bf}\cite{suss} would imply that some of the information
in this spacetime could be viewed in complementary ways by causally
disconnected observers.  Only the observer in the true vacuum would
have (in principle) access to all of the information.  The inhomogeneous,
and false vacuum observers would be viewing very special factor spaces of
the full Hilbert space.  In fact, the appropriate bound on the entropy
appears to follow from Bousso's proof of the N-bound\cite{raphN} for
spherically symmetric spacetimes.  The only thing that confuses me about
this claim is that it is not clear how to identify the quantity that
Bousso calls the cosmological constant in the present models.
The bound proves what we want it to if I identify the cosmological constant
with the vacuum energy at the true minimum.  I do not see {\it a priori}
why I could not have chosen to call the false vacuum energy the cosmological
constant instead.

I should note that in ``practical'' applications the number of true vacuum
states is much much larger than the number available to either of the
other observers.  That is, although experiment and the anthropic principle
force us to accept a fine tuning of parameters to obtain a small
cosmological constant in the low energy effective
Lagrangian which describes our observations, there is no reason to insist
on any other fine tuning.  Thus, the size of the horizons seen by the
inhomogeneous and false vacuum observers is determined by microscopic
parameters.  The probability for forming a false vacuum bubble in the
true vacuum is thus
very small.  Furthermore, the instanton which describes the false vacuum bubble
formation, actually describes the creation of a spacetime which is, over
most of its volume, a small excitation in the basin of attraction of the
true vacuum.   According to Cosmological Complementarity, there are many
different observers in this spacetime, but most of them observe the instanton
event as just
a small excitation of the true vacuum.  A small subset of rather short lived
observers can actually see either the false vacuum, or the inhomogeneous state
created.   Furthermore, the waiting time for a true vacuum observer to
see the instanton event, is exponentially larger than the time for quantum
fluctuations to destroy the measurements that a true vacuum observer can make
with any instrument whose workings can be described by cutoff quantum field
theory\cite{bfp}. It is also exponentially larger than the expected time for
the observer to be killed by spontaneous nucleation of a black hole at his
position.
 Nonetheless, there is a finite, though small, probability
that the true vacuum observer will see the instanton event which corresponds
to the creation of a region of false vacuum, even though he never
sees the false vacuum itself. An interesting question, to which I do not
know the answer, is whether this observer can create a non-vacuum state
which will relax to the instanton configuration.  Since the region of
false vacuum is to be created outside the cosmological event horizon of
the observer, a negative answer does not automatically follow from the
negative answer provided by Guth and Farhi\cite{gf} to the corresponding
question in flat spacetime.

There is another situation which some authors might think of as
having practical interest.  We can consider a situation in which
our own world is identified with the false vacuum, and there is
another true vacuum with vanishing or much smaller positive
cosmological constant.   Such a system apparently needs at least
two fine tunings for its construction.  Moreover, the second
fine tuning has no anthropic justification.  I consider it unlikely
that any quantum theory of gravity will be found which would be
described by such a low energy Lagrangian.   Nonetheless, since
such constructions have been considered in the literature, it is
worth discussing them.  I will consider only the case where the true
vacuum has a nonzero cosmological constant.

In this case, two classes of observers have macroscopic lifetimes, namely
those observers who see the instanton as creating a fluctuation in the
basin of attraction of either of the vacua.  For a generic lagrangian with
two fine tunings, the inhomogeneous observers will still have a lifetime
of microscopic proportions.   Again, although the waiting time for the
false vacuum observer to see the instanton event is large compared to
any practically realizable lifetime for his observations, there is a tiny
probability that he will observe it.   Most static observers in the original
false vacuum dS space will make a quantum
jump to a small excitation of the true vacuum,
while a few observers will just see a small excitation of their own false
vacuum.  Thus,
there will always be a few heretics who will remain forever justified
in their belief that vacuum decay never occurs.

\subsection{Singularity of the flat space limit}

In an attempt to gain some insight both into the form of the lowest action
instanton, and the issue of negative modes, I have examined a limit in
which the gravitational equations formally approach those of QFT in flat
spacetime.  It turns out that this limit is singular, and generates as
much confusion as it solves.  The requisite limit is one in which $U = U(a P)$
where $a$ is a large parameter.  In terms of dimensionful scalar fields, this
means that the potential has substantial variation on field scales much
smaller than the Planck scale, as one would have expected for standard model
Higgs potentials.  Defining
\eqn{P}{p = a P,}
\eqn{x}{x = a z}
and
\eqn{r}{r = a \rho,}
the equations become
\eqn{flateqa}{p^{\prime\prime} + 3{r^{\prime}\over r}p^{\prime} + U_p = 0}
\eqn{flateqb}{(r^{\prime})^2 = 1 + {r^2 \over 3 a^2} ({{p^{\prime}}^2 \over 2} + U). }
Primes now denote derivatives with respect to $x$.
In the formal $a \rightarrow\infty$ limit, and taking the positive
sign for the square root, these become the flat space equations
and the interval for $x$ becomes half infinite.

Now consider following the flat space instanton, but keeping the ``small''
terms in the equations.  For large $x$ one enters a region of large $r$
with very small $p^{\prime}$.  Eventually, the small term overwhelms the
one in the equation for $r$.  Furthermore it is negative.   Thus, the
perturbation is never negligible.  This is not surprising.  The topology
of the Euclidean manifold suffers a discontinuous change when $a$ goes
to infinity.  The finite $a$ manifolds are always compact.

In fact, we can see that boundary conditions for $P_0$ very close to that of
the flat space instanton do not give rise to smooth compact solutions.
When $r$ gets large, its derivative inevitably changes sign and it begins
to decrease.   But in this region, solutions close to the flat space
instanton are very close to the false vacuum with small velocity, heading
toward the false vacuum.  When anti-friction sets in, they will be driven
over the barrier and never come back.  We will have to choose initial values
at some finite distance ({\it i.e. } not proportional to an inverse power
of $a$) from the flat space value in order to find a solution which does
not fly over the barrier. Furthermore, in most nonsingular solutions in
which $r$ grows to a value of order $a$ before beginning to decrease,
there must be many oscillations.  This is because the $p$ equation does
not contain $a$, so its natural oscillation periods are of order one,
while for $r$ of order $a$ or less, the decrease of $r$ by finite
amounts also takes place over ranges of $x$ of order one.

For large $a$, it appears that the lowest action, nonsingular, instanton
is quite close to the solution which sits at the true vacuum.  As noted above,
the increase and decrease of $r$ must take place over an $x$ range of order
$a$, while the $p$ equation does not contain $a$.  In order to obtain
a nonsingular solution which does not have multiple oscillations, we must
start near a maximum of $U$.  If we start within $1/a$ of the maximum, with
zero velocity, then $p$ will change by order one in an $x$ range of order
$a$.   By tuning the initial distance from the maximum we should be able
to arrange that $p_x$ vanishes at the second zero of $r$.  Note that this
can only happen when $p$ is to the right of the minimum of $U$.  Once
$r$ begins to decrease, and anti-friction sets in the solution can decelerate
only as it climbs the upward slope to the false maximum.

\section{Other instantons}

There are two other types of gravitational tunneling process, which have
been widely discussed in the literature: the membrane creation of Brown
and Teitelboim\cite{bt} and ``decays to nothing'',
as exemplified by the old work
of Witten\cite{wkk}.   The Brown-Teitelboim work is closely related to
that of Coleman-de Luccia.  Decays to nothing are somewhat tangential to
the work of this paper, but our general philosophy throws some light on
them as well, so I will devote a small subsection to them below.

\subsection{Membrane creation}

The Brown-Teitelboim process seems to be a system in which the thin-wall
approximation is nearly exact.  It consists of the nucleation of a spherical
shell of membrane in a background p-form gauge field, and its analysis is
modelled on Schwinger's calculation of pair production in a constant
electric field.   More precisely, it is the higher dimensional analog
of pair production in $1+1$ dimensional electrodynamics - the massive
Schwinger model.

The Schwinger model is amenable to systematic analysis using the technique
of bosonization.  It is exactly equivalent to a scalar field theory with
Lagrangian
\eqn{schwing}{{\cal L} = {1\over 2}[(\nabla\phi )^2 - {e^2 \over \pi} \phi^2
+ M^2 cos(\phi )].}
In this formulation of the theory, pair production is simply described as
vacuum decay of the false vacua at minima of the cosine.  These correspond
to backgrounds with fixed quantized values of electric flux.   The Schwinger
approximation to tunneling, in which one considers a circular Euclidean
electron path, corresponds to a step function approximation for $\phi$, and
is valid in only a limited region of the parameter space
of the model. Even in that regime there are corrections to the
step function approximation to $\phi$ .

In higher dimensions, although there is no exact bosonization formula,
we can still introduce a scalar field through
\eqn{flux}{F_{\mu_1 \ldots \mu_d} = \epsilon_{\mu_1 \ldots \mu_d} \phi}
and write a derivative expansion for the effective action of $\phi$, which
includes effects of virtual membranes.
General considerations show that it will have the above form, except that
the cosine will be replaced by a more general periodic function.
There are other, non periodic, terms coming from higher powers of the
gauge field strength, but they are negligible at low energy.

Thus, I would claim that a more precise analysis of the Brown-Teitelboim
process introduces corrections to the thin wall approximation, and that
one must deal with the issues studied in this paper here as well.

In particular, in the discussions of relaxation of the cosmological
constant by membrane creation\cite{relax}, it is necessary to introduce
a large negative cosmological constant in the low energy effective
field theory in order to obtain vacua with very small cosmological
constant.  Tunneling transitions between a large number of vacua
are then postulated.   Many of these will be between two dS vacua.
Proponents of this scenario would welcome a principle (such as the
one we have proposed) which implied that there could never be transitions
to the negative energy states. Nonetheless,  it is disturbing to have the
whole structure depend on the existence of the negative vacuum energy state,
which can never be accessed by the system. My tentative conclusion
is that such scenarios will not be meaningful in a real theory of
quantum gravity.

\subsection{Decay to nothing}

Decay to nothing is an intrinsically gravitational effect, described
in the simplest case by an instanton which is simply the Euclidean
Schwarzchild solution.   One treats the periodic coordinate as
spatial, and analytically continues one of the coordinates of a sphere.
The resulting manifold is asymptotically a circle times Minkowski space.
In the interior it has nontrivial topology .  This instanton is only
allowed in nonsupersymmetric compactifications, because the fermion
boundary conditions on the circle at infinity, violate supersymmetry.

The colorful phrase `` decay to nothing'' is not really accurate.  In
conventional vacuum decay in Minkowski space, any timelike observer is
hit by a bubble wall in finite proper time, and its trajectory penetrates
into a region of true vacuum.  Assuming the measuring apparatus survives
the passage through the bubble wall, it then measures the properties
of the true vacuum forever after.   In decays to nothing, the trajectories
of timelike observers are bound to the bubble wall, but they cannot
pass through it because there is no spacetime behind it.  Spacetime has
an asymptotic null infinity as well, that ends when it intersects the
bubble wall.

Thus, there is no decay to a static vacuum state, but rather to a
spacetime with no timelike Killing vector.  What should one consider
the candidate for the stable vacuum state into which the static, SUSY violating
Minkowski space decays in this situation?

The answer to this question is intertwined with another serious problem
with all known vacua in which instantons describing decays to nothing
exist.   They are perturbatively unstable.  That is, although the tree
level theory has solutions with a causal boundary similar to that of
Minkowski spacetime\footnote{In the case of the Lorentzian continuation
of the Witten bubble, future null infinity is cut off by its intersection
with the ``bubble of nothing''.   By abuse of language, I call such a
spacetime boundary ``similar to that of Minkowski space''.}.
, at one loop an effective potential is generated which completely
changes the asymptotic behavior of solutions.  In all examples I know
of, the one loop corrected equations of motion have no solutions that
remain under perturbative control.

The effects of loop corrections to the equations of motion are formally
of lower order in the perturbation expansion than the instanton contribution.
Thus, it seems mathematically inconsistent to consider the instanton
as the dominant mode of ``decay'' of the tree level Minkowski vacuum state.
There does not appear to be a metastable Minkowski vacuum state at all.

Equally disturbing is the fact
that the instanton solution gives us no resolution
of the problem posed by the perturbative effective potential.  If the
instanton indeed produced a bubble of a state which was not subject
to the perturbative instability, we might be willing to overlook the fact
that the perturbative potential destroys the asymptotic behavior of
the instanton.  But no such bubble of true vacuum exists.  All observers
in the Lorentzian continuation of the instanton will feel the destabilizing
effects
of the perturbative corrections to the action.

It would appear that there are only two sensible conclusions about systems
of this type.  Either the semiclassical analysis is not an approximation
to {\it any} quantum theory, or it is a bad approximation to a quantum
theory defined with very different asymptotic conditions.  In either case,
the statement that a Minkowski vacuum state decays via a bubble of nothing
would not be a valid description of the physics.

\section{\bf Conclusions}

Since the seminal work of Coleman and De Luccia, instanton methods have
been widely used in the study of quantum gravity.  In this paper I have
tried to argue that the idea of vacuum decay in quantum gravity must be
re-examined with care.  In fact, we have found no situations in which this
notion makes unambiguous sense.

Implicit in the use of instanton methods in quantum mechanics and quantum
field theory, were two propositions whose validity in theories of gravity
is somewhat dubious.  The first is that states with different ground
state energies lie in the same Hilbert space.   The second is that instantons
demonstrate a decay of a metastable false vacuum state into a stable true
vacuum.  I have recently argued that the first proposition is in fact
false\cite{hetero}.   Quantum theories are defined in terms of
their high energy behavior.  Among the high energy excitations of any
theory of quantum gravity in asymptotically flat, or AdS spacetime, are
stable or metastable black holes.
{\it Asymptotic darkness} is the conjecture that these are
in fact the generic high energy state of the theory.  It is manifestly
true in the AdS/CFT correspondence.   Furthermore, the asymptotic
spectrum of black holes in AdS/CFT is sensitive to the value of the
cosmological constant, which is a discrete, tunable parameter.  Thus,
in this context, the idea that different values of the cosmological
constant correspond to different states of the same theory, which
can decay into a stable ground state, is manifestly false.
The asymptotic spectrum of black hole states in asymptotically flat space,
is completely different than that in {\it any} AdS space, so that
instantons connecting such vacua do not seem to make any sense either.

The basic argument is thus that maximally symmetric
spacetimes with nonpositive values of the cosmological constant
cannot decay into AdS space with a more negative value of the cosmological
constant because the former have, by virtue of their existence as
metastable states, black hole excitations which simply cannot exist
in the putative true vacuum.  Note that in conventional quantum theories,
generic
high energy excitations of the true and false vacuum states are identical,
and there is no tunneling barrier for decay of an excitation of one
into an excitation of the other.  This is manifestly untrue for black
holes.

Here I have argued that the CDL analysis of decay into a state with
negative cosmological constant does not really give us evidence for
transitions between flat or AdS vacua, and AdS vacua with lower cosmological
constant.  CDL pointed out that the spacetime inside the ``bubble of true
vacuum'' is in fact a negatively curved FRW universe with a Big Crunch
singularity.  The Lorentzian continuation of the CDL instanton thus resembles
a black hole.   If the false vacuum is asymptotically flat, then the main
difference is that the bubble wall intersects null infinity at finite
affine parameter, and there is no timelike infinity.   One might argue that
such configurations should simply not be allowed, because the quantum theory
is defined only in terms of configurations that are truly asymptotically
flat.  For example, if the quantum theory is formulated as a holographic
theory on all of null infinity, it would be hard to understand how it
could contain configurations like the Lorentzian CDL instanton.

We are familiar in quantum field theory with the statement that field
configurations should satisfy the boundary conditions at infinity.
In field theory the Euclidean and Lorentzian versions of this statement
are equivalent to each other. Fundamentally, I believe that this is
a consequence of the Cauchy-Kowalevska (CK) parametrization of solutions
of the Lorentzian equations.  Requiring spatial fall off of the initial
data is equivalent, upon analytic continuation, to falloff in all directions
in Euclidean space.

It is becoming more and more clear, that in theories of gravity, because
of the ubiquitous occurrence of spacelike singularities, the CK parametrization
of phase space is inadequate.   We do not know a general rule for describing
the global phase space of general relativity, but there are special cases
where we know the answer\footnote{If we accept a version of
the famous Cosmic Censorship conjecture.}.  That is, we know how to parametrize
the space of asymptotically flat or AdS solutions, if we are willing
to accept the conjecture that smooth (perhaps locally bounded) data on
the conformal boundary define solutions all of whose singularities are
shrouded behind black hole horizons.  Technically, asymptotically flat or
AdS asymptotics, includes the provision that the spacetime has a conformal
boundary identical to that of the indicated maximally symmetric spacetime.

The Lorentzian continuations of CDL instantons for decay of flat or AdS space,
do not obey this boundary condition, despite the fact that the Euclidean
solutions fall off at infinity, and have finite action.
The interior of the bubble contains a spacelike singularity and the singularity
intersects null infinity at a finite affine parameter.  Future null infinity
is not geodesically complete in these spacetimes.  Timelike infinity is
a $d-2$ sphere instead of a point.

I believe that the simplest conclusion from all of these data is that
decay of a maximally symmetric space of nonpositive cosmological constant
into another one is not a valid concept in properly formulated theories
of quantum gravity.  The asymptotics at large spacelike distance,
 and the value of the cosmological constant are built into the structure
of the quantum theory.

One may object that by appropriate choice of discrete parameters ({\it e.g.}
fluxes) one can construct effective potentials in string theory that
contain metastable, as well as stable, AdS vacua, in a regime where
perturbative calculations are sensible.   I would conjecture that in all
such cases the CDL instanton will not exist.  We have seen that this is
a generic possibility, which depends on the details of the potential.
Alternatively, the existence of the instanton might be taken to imply that
the metastable state described by the false vacuum did not exist.
I have argued elsewhere that the concept of an effective potential is
suspect in theories of quantum gravity\cite{hetero}.  What is certain
is that the hypothetical instanton is not evidence for decay into the
stable AdS vacuum.

Our assessment of the validity of the CDL analysis for the case of a
false vacuum with positive cosmological constant was much less conclusive
because there does not yet exist a rigorous formulation of the
quantum theory of dS spacetime.  If it is a theory with a finite number
of states, then it is obvious that vacuum decays cannot occur.  This is
not the end of the story.  We can imagine, in a quantum theory of dS
space with a small cosmological constant, a metastable factor
of the Hilbert space, which behaves in some way like a dS space of
larger cosmological constant. Similarly, in the infinite dimensional
Hilbert space of the quantum theory
of an open Big Bang FRW cosmology (the apparent geometry of the interior
of the CDL bubble describing decay to a space of vanishing cosmological
constant), we can imagine a finite tensor factor which describes
a metastable dS space.  It is to these hypothetical situations that
the CDL instanton analysis might apply.

I found that the CDL instanton equations generically have an infinite
discrete set of solutions, with an accumulation point at the Hawking-Moss
instanton.  This is true independent of the parameters in the Lagrangian.
{\it Thus, I disagree with the claims of \cite{mss} that metastable dS minima
of an effective potential can sometimes be exactly stable.}
The Lorentzian continuation of the generic solution describes
two bubbles nucleated in an inhomogeneous spacetime, which is a dS fibration.
The two bubbles are out of causal contact.  Furthermore, many timelike
observers in the spacetime are out of causal contact with either of the
bubbles.  Each bubble evolves to either the false or the true vacuum,
depending on our choice of instanton.   I argued that the lowest action
instanton was probably the one with the smallest number of oscillations,
in which case, one bubble evolves to the true and one to the false vacuum.
The interpretation I proposed of this instanton was thermal.  The system
described by the effective field theory is a quantum system with a number
of states given in terms of the true cosmological constant.  It has a subsystem
with a much smaller number of states (within the range of validity of
the semiclassical approximation, and in the absence of extreme fine tuning)
which local observers view as a dS space with the false value of the
cosmological constant.  Transitions between these two states can occur,
because local observers in either state are in thermal equilibrium with
a random system on their horizon.  The CDL instanton describes both of these
transitions with probabilities that satisfy a law of detailed balance.

The actual effect of instanton transitions on local observers is somewhat
peculiar.  There are, in the aftermath of the instanton event, three classes
of local observers, whom I have characterized as {\it true}, {\it false} and
{\it inhomogeneous}.   In the generic case where only one fine tuning has been
made (to make the true cosmological constant much smaller than its natural
value $M^4$.), only the true observers live in a spacetime with a
macroscopic cosmological horizon.  When the instanton occurs in the true
vacuum, the true observers see it as a small excitation of the vacuum, which
quickly inflates away.  Observers outside the cosmological horizon of any
true observer see a quantum jump to an inhomogeneous static spacetime, or
to a small excitation of the false vacuum.  Note that in this case, both
inhomogeneous and false observers must be microscopically small to fit inside
the cosmological horizons of their respectively regions. If we do two fine
tunings, then the false vacuum observers can also be macroscopically large.

This case is the one in which it is most interesting to think about the effect
of the instanton transition on an observer initially in the false vacuum.
The majority of such observers will, shortly after the nucleation of the
bubble, be hit by the bubble wall.  If these observers survive the collision,
they find themselves in
the true vacuum, able to perform a much larger number of observations than
they had previously thought possible.  But there will always be some heretics,
who see the instanton event as merely a thermal/quantum jump to an excited
state of the false vacuum.  These infidels will live out the rest of their
lives in their narrow world, and can continue to refuse to believe in the
existence of vacuum decay. Fortunately, these fanatics remain forever causally
disconnected from the true believers, and bother their existence no more
than does a quantum fluctuation on the faraway horizon.

The instanton events always have a very small probability, much smaller than
the probability that quantum fluctuations will render an observer's
measuring device inoperable, or that a black hole will be spontaneously
nucleated on top of him.  Nonetheless, quantum mechanics remains quantum
mechanics, and the theory predicts that there is a finite but small
probability that a given observer will actually get to view the decay
of the vacuum.  He should live so long!

The above interpretation of CDL instantons is appropriate for the case
where both true and false vacua have positive cosmological constant.
In the case where the true cosmological constant is zero, other issues
must be addressed.  One must first establish that the negatively curved
FRW universe which plays the role of the true vacuum, is actually a sensible
solution to quantum gravity.   Then one must understand whether processes
in such a spacetime could indeed excite a portion of it into the false
dS minimum of the potential.  The case of a negative value of the
true cosmological constant might also make sense for a false dS vacuum.
At a minimum, one would require the number of states observable in the
Big Crunch universe that results from vacuum decay, to be larger than
the number in the false dS vacuum.   This will be true for a range of
parameters in the effective potential.

To summarize, we have not found any gravitational context in which the
concept of false vacuum decay makes unambiguous sense.   The most promising
case is that of dS to dS transitions.  A model in which this case
is of phenomenological interest would require two fine tunings, one of
which could not be justified even by the anthropic principle.  The case
of dS space tunneling to a space with vanishing cosmological constant
might provide a model of quintessence with the dubious virtue of never
being testable.   Both this case, and the case of dS space decaying into
a spacetime with negative cosmological
constant, might be realizable in perturbative stringy constructions.
The analysis of this paper suggests several constraints that must be
satisfied by such constructions, and one could imagine checking whether
these constraints were satisfied.  The most important message of this
paper is that the use of effective field theory to study these issues
must proceed with caution.  Vacuum decay in the presence of gravity is
a much more subtle issue than even Coleman and DeLuccia realized.  Later,
discussions, which have largely ignored the subtleties, have given
the impression that establishing the existence (or not) of an instanton
solution in the thin wall approximation establishes (or negates) the validity
of the concept of vacuum decay in particular examples.  I would submit that
no such claim has been established. In fact, evidence from rigorous
formulations of quantum gravity in asymptotically AdS space, and from the
principle of Asymptotic Darkness, suggest that the effective potential
formalism which is at the basis of instanton calculations has only
limited applicability in theories of quantum gravity.

Before concluding, let
me caution again that these negative remarks do not necessarily apply to the
discussions of vacuum decay in inflationary cosmology.  One common view
of inflation is that we should describe the entire inflationary patch
as a subsystem of a much larger universe.   Applications of the CDL
analysis in such a situation would conform more closely to the original
condensed matter context, in which vacuum decay was merely an idealization
of extensive behavior of finite systems.  Deep questions about the total
number of states in the universe could be reserved for the discussion of
what the quantum theory of gravity, into which the inflationary subsystem
fits, consisted of.   In this context it might make sense to truncate the
dS instanton and consider only the true vacuum bubble.  Inflationary
cosmology is only locally the same as dS space.



  %




\end{document}